\documentstyle[camera,prbplug,psfig]{article}
\newcommand{\csvo}{Ca_{1-x}Sr_xVO_3}
\newcommand{\cvo}{CaVO_3}
\newcommand{\svo}{SrVO_3}
\newcommand{\Ca}{Ca$^{2+}$}
\newcommand{\Sr}{Sr$^{2+}$}
\newcommand{\ef}{$E_F$}
\newcommand{\degc}{\,$^\circ$C}
\newcommand{\mb}{$m_b$}
\newcommand{\meff}{$m^{\ast}$}
\newcommand{\adir}{$[100]$}
\newcommand{\abdir}{$[110]$}
\newcommand{\ie}{{\it i.e.}}
\draft
\title{Band-width control in a perovskite-type 3$d^{1}$ 
       correlated metal \csvo\@. I\@.\\
       Evolution of the electronic properties and effective mass.}
\author{I. H. Inoue\cite{a.inoue}~\cite{a.etl}}
\address{Precursory Research for Embryonic Science and Technology, 
         Japan Science and Technology Corporation,\\
         and  Electrotechnical Laboratory, 1-1-4 Umezono, Tsukuba 305, 
         Japan}
\author{O. Goto}
\address{Department of Electronics and Communications, Meiji 
         University, 1-1-1 Higashi-Mita, Tama-ku, Kawasaki 214, Japan}
\author{H. Makino}
\address{Institute of Applied Physics, University of Tsukuba, 1-1-1 
         Tennodai, Tsukuba 305, Japan}
\author{N. E. Hussey\cite{a.hussey}}
\address{Interdisciplinary Research Centre in Superconductivity, 
         University of Cambridge, Madingley Road, Cambridge CB3 0HE, 
         United Kingdom}
\author{M. Ishikawa}
\address{Institute for Solid State Physics, University of Tokyo, 
         7-22-1 Roppongi, Minato-ku, Tokyo 106, Japan}
\date{December 25, 1997}
\begin{document}
\maketitle
%
%
\begin{abstract}
Single crystals of the perovskite-type $3d^{1}$ metallic alloy system 
\csvo\ were synthesized in order to investigate metallic properties 
near the Mott transition.
The substitution of a \Ca\ ion for a \Sr\ ion reduces the band width 
$W$ due to a buckling of the V-O-V bond angle from $\sim\!180^\circ$ 
for \svo\ to $\sim\!160^\circ$ for \cvo\@.
Thus, the value of $W$ can be systematically controlled without 
changing the number of electrons making \csvo: one of the most ideal 
systems for studying band-width effects.
The Sommerfeld-Wilson's ratio ($\simeq\!2$), the Kadowaki-Woods ratio 
(in the same region as heavy Fermion systems), and a large $T^{2}$ 
term in the electric resistivity, even at $300$\,K, substantiate a 
large electron correlation in this system, though the effective mass, 
obtained by thermodynamic and magnetic measurements, shows only a 
systematic but moderate increase in going from \svo\ to \cvo, in 
contrast to the critical enhancement expected from the Brinkmann-Rice 
picture.
It is proposed that the metallic properties observed in this system 
near the Mott transition can be explained by considering the effect of 
a non-local electron correlation.
\end{abstract}
\pacs{71.28.+d, 71.30.+h, 71.20.Be}
%
%
%
\section{INTRODUCTION}
Despite extensive investigations on 3$d$ transition-metal (TM) 
oxides,\cite{reviewTM} there remain many more mysteries still to 
unravel.
The discoveries of metal-to-insulator transitions (MIT) in $3d$ TM 
oxides with a partially filled $3d$ band, for example, has given us 
great incentive to re-examine several previous studies of the 
electronic states in these TM oxides.

The most important feature of this kind of 3$d$ TM oxides is that 
simple one-electron band theory is no longer sufficient to give a good 
account of the electronic states, since the electron correlations are 
much larger than expected for the one-electron 
band-width.\cite{peierls}
Mott first introduced the concept of MIT caused by a strong Coulomb 
repulsion of electrons.\cite{mott}
Although the description of the MIT (Mott transition) is still argued 
from various points of view,\cite{review} a more challenging problem 
lies in the metallic phase near the Mott transition, where a 
narrow-band system is known to show {\it anomalous} metallic 
properties, and substantial enhancement of the fluctuations of spin, 
charge and orbital correlations is observed.
This problem has been investigated with renewed vigor since the 
discovery of high-T$_{\rm c}$ cuprate superconductors and although a 
number of enlightening works have been done so far, still we cannot 
grasp a comprehensive view of the whole physics.

As one of the open questions, in this paper, we focus on the problem 
of the effective mass in the perovskite-type 3$d^{1}$ correlated metal 
\csvo.
An important manifestation of the mass enhancement in the 
perovskite-type light-$3d$ TM oxides has been given by Tokura {\it et 
al}.\cite{TokuraLSTO}
They reported filling-dependent electronic properties in the 
Sr_{1-x}La_xTiO_3 system near the MI transition around $x=1$\@.
The LaTiO_3 ($x=1$) material behaves as an insulator below 300\,K and 
antiferromagnetic ordering of Ti $S=1/2$ spins occurs at $T_N = 120 
\sim 150$\,K\@.\cite{LTOMI}
They also reported that Fermi-liquid-like behavior was observed even 
in the immediate vicinity of the MI phase boundary, with a critical 
increase of \meff\ arising from the effect of the enhanced electron 
correlations.
Within the framework of Fermi-liquid theory, the only way to approach 
MIT continuously is to realize the divergence of the 
single-quasi-particle mass \meff\ at the MIT point.\cite{FLMIT}
The critical behaviors observed in the Sr_{1-x}La_xTiO_3 system are 
fairly systematic, thus provoking intense theoretical study; however 
there is still room for arguments, especially in the following points:
\begin{enumerate}
    \item Tokura {\it et al}.\ compared the effective mass \meff\ to the 
    free electron mass $m_{\rm o}$\@.  However, the $x$-dependence of the 
    ``band-mass'' \mb\ should also be taken into account.
	Compared with the value of \mb\ for the similar system \cvo, 
	the mass enhancement of the Sr_{1-x}La_xTiO_3 system is not so large, 
	except for the region $x > 0.95$\@. 
	\item  The critical increase of the value of \meff\ in the 
	Sr_{1-x}La_xTiO_3 system is only seen in the region very close to 
	MI transition boundary\cite{LSTOnew}.
	However, in this region,
	it is not obvious whether Fermi-liquid theory is still valid.
	In fact, in the region of significant mass-enhancement $(x > 0.95)$,  
    the number of carriers seems to be depleted.\cite{LSTOnew} 
	\item  Another filling-dependent MI transition is observed 
	in  the Y_{1-x}Ca_xTiO_3 system.\cite{YCTOtokura,YCTOiga} However 
	MIT occurs around $x=0.4$ which is relatively far from 
	integral filling. 
    Nevertheless, the effective mass shows a conspicuous enhancement in 
    the vicinity of MIT similar to that seen in the 
    Sr_{1-x}La_xTiO_3 system.
	Thus, it seems reasonable to suppose 
	that this kind of mass-enhancement, observed in those 
	``filling-control'' systems close to MIT, 
	might be induced by fluctuations or inhomogeneity of the insulating 
	phase near the boundary of MIT.
\end{enumerate}
The above problems can be due to the fact that the critical behaviors 
depend on a path along which a system approaches the boundary of 
MIT\@.  In the Sr_{1-x}La_xTiO_3 system, the band-filling is 
dominantly controlled instead of the band-width.

The question then arises; how does the effective mass in the metallic 
state actually change as we change solely the electron correlation 
without changing the band-filling?
In order to elucidate this issue, another type of systematic 
experiment is required; \ie, we need to control only the 3$d$ band 
width $W$ in a particular system while keeping the number of carriers 
fixed.

Representative examples are the pressure-induced MIT reported in 
V_{2}O_{3},\cite{VTwoOThree} where hydrostatic pressure modifies 
$W$\@.
However, for a quantitative discussion, we need to know the change of 
the lattice constants under pressure. Moreover, in general, the 
anisotropic compressibility due to the anisotropy of the lattice 
structure affects $W$ in a complex manner.
Other examples are found in nickel-based compounds: 
the perovskite-type {\it R}NiO_{3} with {\it R} 
of the trivalent rare-earth ions (La to 
Lu),\cite{RNiOThree} and the pyrite-type chalcogenide system 
NiS_{2-x}Se_{x}\@.\cite{NiSTwo}
The insulating state of these nickel compounds is classified as a 
charge-transfer insulator rather than a Mott-Hubbard insulator 
in the so-called Zaanen-Sawatzky-Allen classification scheme of TM 
compounds.\cite{ZSA,hufner} 
Therefore, MIT occurs as a 
closing of the charge-transfer gap with increase of the $p$-$d$ 
hybridization.
Thus, it is inevitable that MIT is not described by the simple model 
of the Mott transition and the metallic state is more complicated.

Based on these considerations, we have synthesized a solid solution of 
the perovskite-type metallic vanadates, \cvo\ and \svo, in order to 
investigate the metallic state near the Mott transition more simply 
with a systematic band-width control.  We have succeeded in obtaining 
single crystals of the homogeneous metallic alloy system \csvo\ with 
nominally one $3d$ electron per vanadium ion.
In the \csvo\ system, as we isovalently substitute a \Ca\ ion for a \Sr\ 
ion, a lattice distortion occurs.
This is governed by the so-called tolerance factor $f$ of the 
perovskite-type compounds ABO_{3} 
defined as
\[
f=\frac{R_{\rm A}+R_{\rm O}}{\sqrt{2}\left(R_{\rm B}+R_{\rm 
O}\right)}\;,
\]
where $R_{\rm A}$, $R_{\rm O}$, and $R_{\rm B}$ are the ionic radii
of the A ion, the O ion (oxygen), and the B ion, respectively.
When the value of $f$ is almost 1, the system is cubic; while for 
$f < 1$, the lattice structure changes to rhombohedral and then to the 
orthorhombic GdFeO_{3} type.
In the GdFeO_{3} structure, it is known that the B-O-B bond angle 
decreases continuously with decreasing 
$f$ almost irrespective of the set of A and B\@.\cite{tolerance}
According to the literature,\cite{shannon} the ionic radii of \Ca, 
\Sr, V^{4+}, and O^{2-} ions are 1.34, 1.44, 0.58, and 1.40\,{\AA}, 
respectively.
Thus we obtain a value of $f$ of 1.014 for \svo, and 0.979 
for \cvo, corresponding to a V-O-V bond angle of 
$\sim\!180^\circ$ for \svo\ and $\sim\!160^\circ$ for \cvo\@.
The buckling of the V-O-V bond angle reduces the one-electron 
$3d$-band width $W$, since the effective $3d$-electron transfer 
interaction between the neighboring V-sites is governed by the 
supertransfer process via the O $2p$ state.

Thus, the ratio of the electron correlation $U$ normalized to $W$ ($U$ 
is considered to be kept almost constant by the substitution) can be 
systematically controlled in \csvo\ without varying the nominal 
carrier concentration.
Furthermore, the V-O-V bond angle of \cvo\ $(\sim\!160^\circ)$ is 
almost equal to insulating LaTiO_3, so it is reasonable to 
consider that \cvo\ is close to the MIT boundary and thus is an 
ideal system for the investigation of the metallic state near the 
Mott transition.\cite{commentOtherPossibility}
In fact, some spectroscopic manifestation of the strong electron 
correlation has been reported 
already,\cite{inouePRL,morikawa,marceloPRLCSVO,makinoCSVO} showing 
that there is significant 
spectral weight redistribution in the \csvo\ system.
Therefore, the effective mass of this system, especially at the $x=0$ 
end (\cvo), is expected to be enhanced as discussed for 
Sr_{1-x}La_{x}TiO_{3} near the insulating composition LaTiO_3\@.

Nevertheless, the \csvo\ system does {\it not} show 
such a significant enhancement of the effective mass.
The goal of this paper is to reveal intriguing behavior in the 
evolution of the effective mass, as we control the $U/W$ ratio in this 
system.
Details of the experiments, especially the method of preparing single 
crystals of this new vanadate system, are described in 
Sec.~\ref{sec.experiments}.
We discuss the cubic-orthorhombic lattice distortion 
in Sec.~\ref{sec.lattice}\@. 
The results from magnetic susceptibility measurements and the obtained 
effective mass \meff\ are shown in 
Sec.~\ref{sec.magnetaization}, and compared to 
\meff\ deduced from the electronic specific heat coefficient in 
Sec.~\ref{sec.gamma}\@.
The Sommerfeld-Wilson's ratio is found to be almost equal to 2, that is 
strong evidence of the large electron correlation.
The electric resistivity data are analyzed by a model incorporating 
the electron-electron interaction ($T^{2}$ term) as well as the 
electron-phonon interaction (Bloch-Gr\"{u}neisen term) in 
Sec.~\ref{sec.resistivity}\@.
It is noted that the Kadowaki-Woods ratio lies in the same 
region as the heavy Fermion compounds.
Finally, we discuss the effect of non-local electron correlations, \ie, 
the momentum-dependent self-energy, which can be significant near the 
Mott transition, in order to explain consistently both the strong 
electron-correlations and the missing enhancement of the effective mass.

\section{EXPERIMENTAL}
\label{sec.experiments}
A `ceramic method' was employed in order to prepare poly-crystalline 
samples.
4N CaCO_3, SrCO_3 and VO_2 were used 
as starting reagents.
We prepared CaO and dried SrCO_3 by pre-heating both the CaCO_3 
and SrCO_3 compounds in air for 24 hours at 1000\degc, and weighed 
the powders while they were still over 100\degc\@.
We confirmed that the dried CaO and SrCO_3, as well as VO_2, were all 
single phase by x-ray diffraction (XRD).
The starting compounds, CaO, SrCO_3, and VO_2 were then mixed in the 
required molar ratio %
\( {\rm Ca }: {\rm Sr} : {\rm V}= 1-x : x : 1 \) %
and then calcined several times at $1250$\,\degc\ in 
flowing argon atmosphere ($\sim$\,1000\,cc/min) with  
intermittent grindings.
Because the reaction proceeds in the solid state, the reaction rate 
depends on the diffusion rate of the constituents through the product 
phases.
As the reaction proceeds, diffusion paths become longer, and hence 
the reaction rate decreases.
Therefore the intermittent mechanical grinding of the reaction 
product is important in this method.

As the Sr concentration is increased, it is required to add 
hydrogen-gas at a rate up to $\sim$\,50\,cc/min.  
The amount of hydrogen-gas flow for each calcination process must be 
controlled in order to avoid too much reduction and peroxidization.
The amount of the oxidation was conveniently checked by examining the 
XRD spectrum, \ie\ the lattice constants, of the reaction product 
every time after the intermittent grinding. 
This process was repeated until completion of the reaction.

Finally, the powder was put into rubber tubes and each tube was pressed 
under hydrostatic pressure of $1000$\,atm to form a cylindrical rod of 
$\sim$\,6\,mm diameter and $\sim$\,10\,cm length.
The rods were sintered at 1300\degc\ in the same atmosphere 
described above.

Single crystals of \csvo\ were grown by the floating zone (FZ) method 
in an infrared-radiation furnace (Type SC-N35HD, Nichiden Machinery 
Ltd.) with two $1.5$\,kW halogen lamps as radiation sources.
At first, the sintered rod is cut into two parts: one is $\sim$\,2\,cm 
long for the ``seed'' rod, which is held at the top of the lower-shaft, 
and the rest of the sintered rod, called the ``feed'' rod, is 
suspended at the bottom of the upper-shaft.
%
%
Each rod is rotated at $\sim$\,20\,rpm in opposite directions.
The lamp power is raised gradually until both the rods are melted, 
then the molten zone is attached to the top of the seed.
The molten zone is passed through the whole feed rod at a rate of 
$\sim$\,1\,cm/hour in flowing argon atmosphere without any 
interruption or change of lamp power.
The most important point here is to control the reduction atmosphere 
delicately, depending on the amount of the Sr content $x$, and also on 
the oxygen stoichiometry of the feed rod.
As we increase the value of $x$, it is necessary to add $<\!0.1$\,\% 
hydrogen to the flowing argon.
As soon as this small amount of hydrogen is added, however, the melting 
temperature rises drastically, and the molten zone shrinks unless we 
increase the lamp power once more.
On the other hand, if the amount of hydrogen gas is not sufficient, 
the liquid phase in the molten zone loses viscosity and spills by 
degrees along the rod.
Furthermore, a different phase, which might be a peroxidized phase, 
appears in the molten zone and precipitates on the surface of the zone 
to form an ``antler''.  Thus, a delicate feed-back control 
of the amount of hydrogen gas and the power of the halogen lamp is 
necessary to obtain a single crystal with sufficient quality.
Typical dimensions of a single-crystalline grain in the resultant rods 
are $\sim$\,$2\times1\times1$\,mm^3\@.

Each crystal was examined by 
powder XRD and by Laue photography to check for homogeneity.
Results are summarized in the next section.
 
The oxygen off-stoichiometry in \csvo\ was determined using a 
Perkin-Elmer TGA-7 thermogravimetric (TG) analyzer from the weight 
gain on heating the sample to around $1300$\,K in flowing air and 
assuming that the final oxidation state of a vanadium ion was $+5$\@.
Neither weight gain due to peroxidation, nor weight loss due to 
desorption of the oxygen was observed, once the highest-oxidized 
material (Ca_{1-x}Sr_x)_2V_2O_7
was obtained.

As-prepared samples contain a certain amount of oxygen defects.  
The result of the TG measurements indicated that, with increasing 
temperature, the samples are abruptly oxidized at around 
$420$\,K\@.\cite{cvoTGA}
Moreover, after this oxidation, the oxygen concentration of the 
samples becomes stoichiometric and no further oxidation occurs until 
the temperature reaches around $700$\,K\@.
Therefore, we were able to prepare samples without any oxygen 
off-stoichiometry by annealing the samples in air at 
$\sim\!200$\degc\ for around $24$\,hours.

The stoichiometry of the ratio \( {\rm Ca }: {\rm Sr} : {\rm V}= 1-x : 
x : 1 \) was confirmed by an inductively coupled plasma atomic 
emission spectrometer (SEICO, SPS7000).  The amount of 
off-stoichiometry in the single-crystalline samples was within the 
error bar, \ie, less than 1\%\@.
  
In order to perform the dc-electric resistivity measurement, the 
single crystalline samples were cut and shaped into rectangular 
parallelepipeds without particular attention to the alignment of the 
crystal axes.
A typical example of the dimension of the parallelepiped was 
$2\times0.5\times0.3$\,mm^3\@.
We also prepared samples with two different alignments: for one 
set of samples, the longest-edge, along which the measuring 
current flows, is parallel to the \adir-axis of the pseudocubic 
perovskite, and for the other set, the current flows along the 
\abdir-axis.
Some of these samples showed clear dependence on the alignment. 
However, this was not due to any features in the electronic structure as 
discussed in the following section.
All the electric resistivity measurements were done with a standard dc 
four-terminal method.
Four copper leads $(50\,\mu{\rm m}\phi)$ were attached with silver 
paste (Du Pont $4922$)\@.
The measuring current was typically $\pm 15$\,mA supplied by a 
constant current source.
Since the $x=1$ sample was much 
smaller than the other ones, we measured its resistivity using the Van 
der Pauw technique in which one places the contacts on the corners and 
rotates the current and voltage configuration.
The data were collected on both heating and cooling cycles.

\mbox{dc-susceptibility} measurements were performed using a 
commercial rf-SQUID magnetometer (Quantum Design, MPMS-II) 
without particular attention to the 
alignment of the crystal axes.
The measuring field was calibrated up to 5\,Tesla with a Pd standard.  

Specific heat data were obtained on polycrystalline 
samples\cite{gammasingle} between $\sim$\,0.5 and $\sim$\,20\,K using 
a semi-adi\-a\-bat\-ic heat-pulse method.\cite{ishikawa}

\section{RESULTS AND DISCUSSION}
\label{sec.results}
\subsection{Lattice constants}
\label{sec.lattice}
\begin{figure}[hb]
  \centerline{\psfig{file=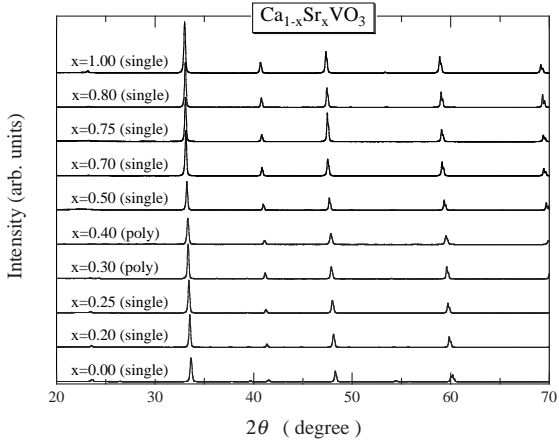,width=7.5cm}}
  \caption{X-ray powder diffraction patterns of single-crystalline \csvo\
           at room temperature. $x=0.3$ and $x=0.4$ samples are poly 
           crystals.}
  \label{f.CSVOXRD}
\end{figure}

\begin{figure}[ht]
  \centerline{\psfig{file=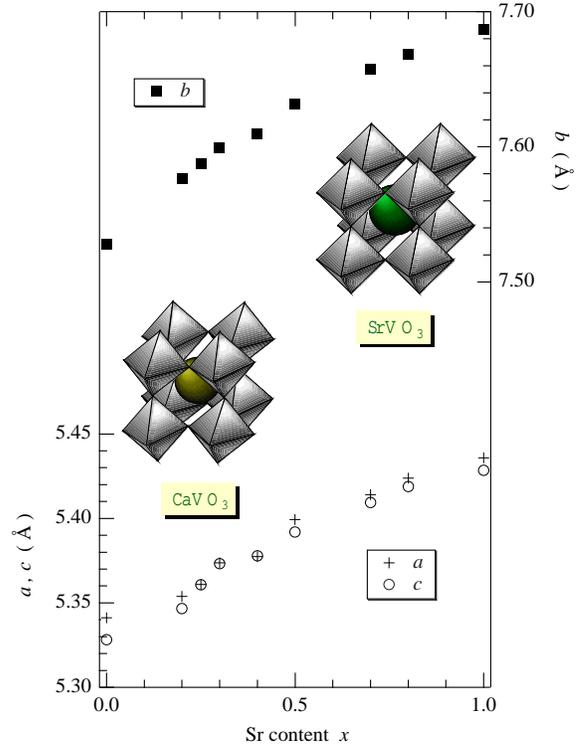,width=7.5cm}}
  \caption{The lattice parameters $a$, $b$, and $c$ of 
           \csvo\ at room temperature, estimated 
           from the XRD patterns. The data 
           are plotted against Sr content $x$\@. Note that each 
           deduced lattice parameter contains around 
           $\pm 0.2$\,\% error. Thus it is not appropriate to discuss
           the exact crystal symmetry based only on this plot.}
  \label{f.CSVOLattice}
\end{figure}

The XRD patterns of \csvo\ for varying Sr content are displayed in 
Fig.~\ref{f.CSVOXRD} and the deduced lattice parameters are shown in 
Fig.~\ref{f.CSVOLattice}.
The lattice parameters change systematically from the orthorhombic 
\cvo\ to \svo\ which is simple-cubic within the error bar 
($\sim\pm0.2$\,\%)\@.
According to the four-axes XRD 
measurement,\cite{cvolatticedata} the lattice constants of \cvo\ are 
$a=0.53185(8)$\,nm, $b=0.7543(2)$\,nm, and $c=0.53433(8)$\,nm.
%
The V-O-V bond-angle is $154.(3)$^{\circ} for V ions on the ac-plane
(V-O bond length is $0.1891(6)$\,nm) and 
171.(0)^{\circ} for V ions along the b-axis
(V-O bond lengths are $0.190(0)$\,nm 
and $0.196(5)$\,nm)\@.\cite{cvolatticedata}
This large buckling of the V-O-V bond angle ($\sim\!160$^{\circ} in average)
is considered to make the 
one-electron $3d$ band-width $W$ of this system smaller than 
that of \svo, where the V-O-V bond angle is almost exactly 180^{\circ}\@.

\subsection{Magnetization}
\label{sec.magnetaization}
\begin{figure}
  \centerline{\psfig{file=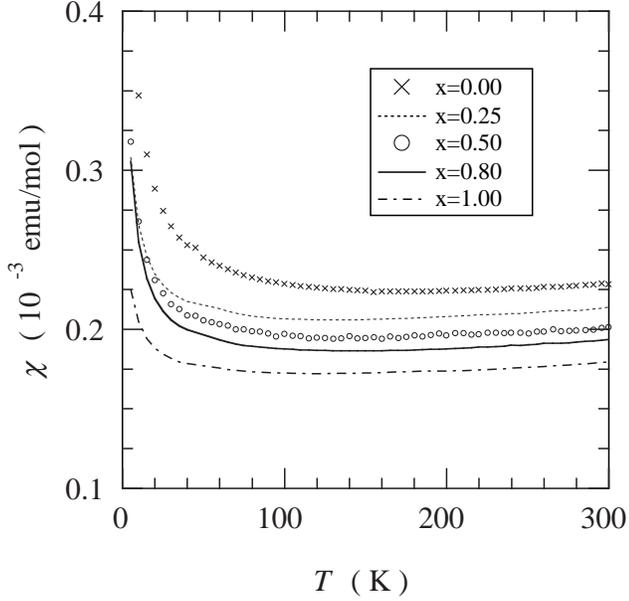,width=\columnwidth}}
  \caption{%
        Magnetic susceptibilities $\chi$ of single-crystalline \csvo\ for 
        $x=0.00,0.25,0.50,0.80,\:\mbox{and}\:1.00$\ 
        measured at 5\,Tesla$\equiv$50000\,Oe
        plotted against temperature $T$\@. Samples were        
        cooled to 4\,K with no applied field and then warmed up to
        300\,K in 5\,Tesla. No significant hysteresis was observed 
        during the heating and cooling cycles.}
  \label{f.CSVOchiAll}%
\end{figure}

Figure \ref{f.CSVOchiAll} shows the temperature-dependence of the 
magnetic susceptibilities $\chi$ of \csvo\ at 
5\,Tesla\,$\equiv$\,50000\,Oe.
Since none of the samples showed any significant hysteresis between the 
heating and cooling cycles, we have plotted data for the 
heating process only.
\begin{figure}[hb]
  \centerline{\psfig{file=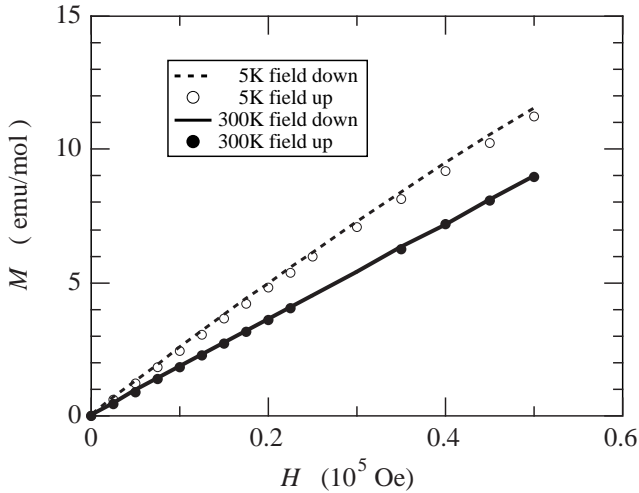,width=\columnwidth}}
  \caption{%
       Magnetization $M$ of \svo\ plotted against the applied field $H$.
       The data were collected while increasing and decreasing the applied field  
       between 0 to 5\,Tesla at both 5\,K and 300\,K\@.}
  \label{f.CSVOMHAll}%
\end{figure}

The field-dependence of the magnetization $M$ of \svo\ is plotted in 
Fig.~\ref{f.CSVOMHAll}\@.
We measured $M$ up to 5\,Tesla while increasing and decreasing the 
applied field $H$ at both 5\,K and 300\,K\@.
At 300\,K, the magnetization curve shows no hysteresis and $M$ depends 
linearly on $H$\@.
This means that only paramagnetic moments contribute to the total 
magnetization.
When we decrease the temperature to 5\,K, the magnetization curves 
become hysteretic and also show a slight upturn.

Here we note that in perovskite-type oxides with the formula ABO_3, 
oxygen and A/B stoichiometries are fairly unstable.
The off-stoichiometry is not accidental but characteristic of these 
compounds.
Even though one tries to obtain sufficiently stoichiometric compounds in 
ABO_3 materials within experimental requirements, the 
off-stoichiometry is still present intrinsically and such inevitable 
defects are called ``native-defects'' to indicate that their 
properties are reproducible.\cite{PhillipsDefect}
We have so far reported the effects of the (unavoidable) oxygen 
off-stoichiometry in \cvo\ by intentionally introducing the oxygen 
defects in varying degrees.\cite{cvoTGA,FukushimaCVO,ShirakawaCVO}
In this study, although we tried to prepare 
oxygen-stoichiometric samples using a delicate annealing 
procedure, there still exists a very small 
but irreducible amount of inevitable oxygen 
defects.

Local moments due to these oxygen defects contribute to 
$M$ as a spontaneous magnetization or in this case a sublattice 
magnetization, because the Weiss temperature is negative.
The value of $M$ is very small compared with that observed in 
CaVO_{2.8}\cite{ShirakawaCVO} and is consistent with the number of local 
moments deduced from the Curie constants ({\it vide infra})\@.

All the magnetic susceptibility data $\chi$ are well reproduced by the 
following formula:
\begin{equation}
  \chi_P + \chi_{dia} + \chi_{core}%
  + \chi_{orb} + \frac{C}{T-\vartheta} + \alpha T^{2}\:,%
 \label{eq.CSVOchiAll}
\end{equation}
where $\chi_P$ is the Pauli paramagnetic term, $\chi_{dia}$ is the 
Landau diamagnetization, $\chi_{core}$ comes from the diamagnetic 
contribution of the core-levels, and $\chi_{orb}$ is due to the 
orbital Van Vleck paramagnetization.
The Curie-Weiss term is attributed to impurities 
such as the native oxygen defects.
The last term is considered to originate from the higher-order 
temperature-dependent term in the Pauli paramagnetism, that is 
neglected in the zeroth order 
approximation, and reflects the 
shape of the density of states (DOS) $D(\omega)$\,Ryd^{-1}/ 
formula-unit
at the Fermi energy \ef\@. 
%

The first two terms are re-written using the effective mass \meff\ and 
the bare band mass \mb\ of the system
[we employ \mb\ deduced from the 
band calculation using a local density approximation 
(LDA),\cite{CVOBandCalc} rather than \mb\ of the non-interacting 
Bloch electrons]:
\begin{eqnarray}
  \chi_{spin}&=&\chi_P+\chi_{dia} \nonumber  \\ %
             &=&\left(\frac{m^\ast}{m_b}-%
                      \frac{m_b}{3m^\ast}\right)\chi_P^{LDA}\:,
  \label{eq.CSVOchiSpin}
\end{eqnarray}
where $\chi_P^{LDA}$ stands for the Pauli paramagnetic term deduced 
from the LDA band calculation:
\begin{eqnarray*}
\lefteqn{\chi_P^{LDA}\:\mbox{[emu/mol$\equiv$erg Oe^{-2}mol^{-1}]}} \\ %
         &=& N\mu_B^2D(E_F) \\ %
		 &=& 2.376 \times 10^{-6} \times D(E_F) \:,
\end{eqnarray*}
where $N$\,mol^{-1} is the number of itinerant electrons per one mole of 
unit formula and $\mu_B=9.274\times10^{-21}$\,erg 
Oe^{-1}\@.\cite{DefOfDOS}

For the third term $\chi_{core}$, we have used the values given in the 
literature,\cite{VanVleck} as summarized in 
Table~\ref{t.CSVOCoreDia}\@.
\begin{table}[ht]
  \centering
  \caption{%
       Core diamagnetism of the constituent ions in \csvo\@.\protect\cite{VanVleck}
  \label{t.CSVOCoreDia}}
  \begin{tabular}{cc}%
   \hline%
      ion         & $\chi_{dia}$\,emu/mol    \\ %
   \hline\hline%
      \Ca\        &  $-13.3\times10^{-6}$    \\ %
   \hline%
      \Sr\        &  $-28.0\times10^{-6}$    \\ %
   \hline%
      V^{5+}      &  $-7.7\times10^{-6}$     \\ %
   \hline%
      O^{2-}      &  $-12.6\times10^{-6}$    \\ %
   \hline%
  \end{tabular}
\end{table}
%
For the fourth term,
we used an $x$-independent value~\cite{VanVleckComment}
of $\chi_{orb}=6.5\times10^{-5}$\,emu/mol 
estimated in another $3d^1$ metallic vanadate system 
VO_2\@.\cite{VO2VV}
Hence, we can fit Eq.~\ref{eq.CSVOchiAll} to 
the observed data.

The obtained values of the Curie-Weiss term $C$ are as small as 
$0.5\!\sim\!2.2\times10^{-3}$\,emu K/mol, and the Weiss temperatures 
$\theta=-6.0\sim\-1.6$\,K, indicating very weak antiferromagnetic 
interaction among the local moments.
These small values of the Curie-Weiss term are considered to be due to 
the V^{3+} $(S=1)$ local impurity moment arising from the ``native 
oxygen defects''.
From the value of $C=2.2186\times10^{-3}$ emu K/mol for \cvo, we can 
evaluate that only 0.22\,\% of the V sites have the $S=1$ local 
moment; this amount of local impurities is inevitable in ABO_3 
materials, but irrelevant for our discussion of the metallic 
properties, such as the value of the effective mass.
The obtained values of the coefficient $\alpha$ of the last term in 
Eq.~\ref{eq.CSVOchiAll} are also very small 
($0.5\sim2.5\times10^{-10}$\,emu K^{-2}mol^{-1}),
implying that the deviation from a temperature-independent Pauli 
paramagnetism is negligible in the temperature range we measured.
Only when we need to estimate the value of \meff\ from 
$\chi_P$ much more accurately will it be necessary to perform the 
measurement up to higher temperatures.
\begin{figure}[b]
  \centerline{\psfig{file=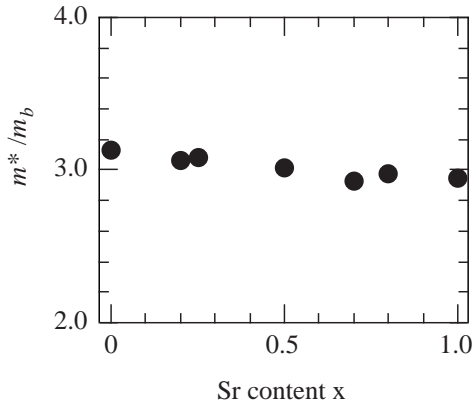,width=6.5cm}}
  \caption{%
       Effective mass \meff\ compared with the LDA band mass \mb\ plotted 
       against Sr content $x$\@.  $m^{\ast}/m_b$ increases systematically 
       in going from \svo\ to \cvo\@. The values are not as large as expected.}
  \label{f.CSVOchiMvsx}
\end{figure}%

The obtained ratio of the effective mass \meff\ to the LDA band mass 
\mb, is summarized in Table~\ref{t.CSVOchi} and also displayed in 
Fig.~\ref{f.CSVOchiMvsx}\@.
\begin{table}
  \centering
  \caption{%
        The effective mass deduced from the fit to the magnetic 
        susceptibility data with Eq.~\protect\ref{eq.CSVOchiAll} and 
        Eq.~\protect\ref{eq.CSVOchiSpin}, where $\chi_{core}$ and 
        $\chi_{orb}$ are fixed to the values in the 
        literature.\protect\cite{VanVleck,VO2VV} $\chi_P^{LDA}$ are 
        calculated from $D(E_F)$ obtained by the LDA band 
        calculation\protect\cite{CVOBandCalc,DefOfDOS} 
  \label{t.CSVOchi}}
  \begin{tabular}{cccc}%
    \hline%
       x  & $\chi_{spin}$\,(emu/mol) & $\chi_P^{LDA}$\,(emu/mol)%
     & $m^\ast/m_b$ \\ %
    \hline\hline%
     0.00 & $2.008\times10^{-4}$     & $6.651\times10^{-5}$%
     & 3.126        \\ %
    \hline%
     0.20 & $1.905\times10^{-4}$     & $6.454\times10^{-5}$%
     & 3.060        \\ %
    \hline%
     0.25 & $1.904\times10^{-4}$     & $6.405\times10^{-5}$%
     & 3.081        \\ %
    \hline%
     0.50 & $1.788\times10^{-4}$     & $6.159\times10^{-5}$%
     & 3.013        \\ %
    \hline%
     0.70 & $1.678\times10^{-4}$     & $5.963\times10^{-5}$%
     & 2.928        \\ %
    \hline%
     0.80 & $1.678\times10^{-4}$     & $5.865\times10^{-5}$%
     & 2.973        \\ %
    \hline%
     1.00 & $1.606\times10^{-4}$     & $5.668\times10^{-5}$%
     & 2.946        \\ %
    \hline%
  \end{tabular}
\end{table}
The value of $m^{\ast}/m_b$ is almost equal to $3.1$ though increases 
gradually and systematically as we decrease the Sr content $x$\@.

As already described, the large buckling of the V-O-V bond angle in 
going from \svo\ to \cvo\ (in \cvo\ $\angle\mbox{V-O-V}\!\sim\!160^\circ$, 
which is almost equal to the insulating 3$d^{1}$ system LaTiO_3) 
can lead this system closer to the boundary of the MI 
transition.
Actually, a significant spectral weight redistribution, 
which is a manifestation of a strong electron correlation, has been 
observed already in this 
system.\cite{inouePRL,morikawa,marceloPRLCSVO,makinoCSVO}
Nevertheless, no significant amount of mass-enhancement can be 
deduced from this magnetic measurement. 
This surprising result motivates us to reconsider whether this system 
may indeed be a correlated metallic system.
However, as well as the reported spectral weight redistribution, the 
measurement of the electronic specific heat and electric resistivity 
described below give us further evidence of strong correlations in 
this system.

\subsection{Electronic specific heat coefficient}
\label{sec.gamma}
An alternative method to evaluate \meff\ is to measure the electronic 
contribution to the specific heat, 
$\gamma T$, 
which
reflects DOS at \ef\@.
$\gamma$ is called the electronic specific heat coefficient.
Using $D(E_F)$\,Ryd^{-1}/formula-unit obtained by the 
LDA band calculation,\cite{DefOfDOS} we can deduce 
the value of electronic specific 
heat coefficient in the non-interacting limit $\gamma^{LDA}$:
\[
\gamma^{LDA} = \frac{\pi^2}{3}k_B^2 N D(E_F)\:,
\]
where $N$ is the number of itinerant electrons per mole in the unit 
formula. 
Then, 
\[
\gamma^{LDA}\:\mbox{[mJ mol^{-1} K^{-2}]}\:= 0.173238 \times D(E_F)\:.
\]
The ratio of the effective mass to the band mass $(m^\ast/m_b)$ is 
deduced from the ratio of the observed $\gamma$ to the calculated 
$\gamma^{LDA}$\@.

Sufficiently below the Debye temperature $\Theta$, the constant 
volume specific heat $C_v/T$ can be 
plotted against $T^2$, \ie, 
\begin{equation}
 C_v/T = \gamma  + \beta T^2 
\label{eq.gammafit}
\end{equation}
in order to separate out the contribution of the ionic degrees of 
freedom $(\beta T^{3})$ dominant at high temperatures.
The coefficient $\beta$ is related to $\Theta$ as follows:
\begin{eqnarray*}
\beta & = & \frac{9 N k_B}{\Theta^3}
 \int_{0}^{\Theta/T}\frac{{\rm e}^z z^4 dz}{({\rm e}^z - 1)^2} \\
&\simeq& \frac{12 \pi^4 N k_B}{5 \Theta^3}\: (T \ll \Theta)\:.
 \end{eqnarray*} 

Here we note that experiments measure the specific heat at constant 
pressure, $C_p$, but we normally compare this result to $C_v$, 
since these two are almost identical in a solid.

\begin{figure}[hb]
  \centerline{\psfig{file=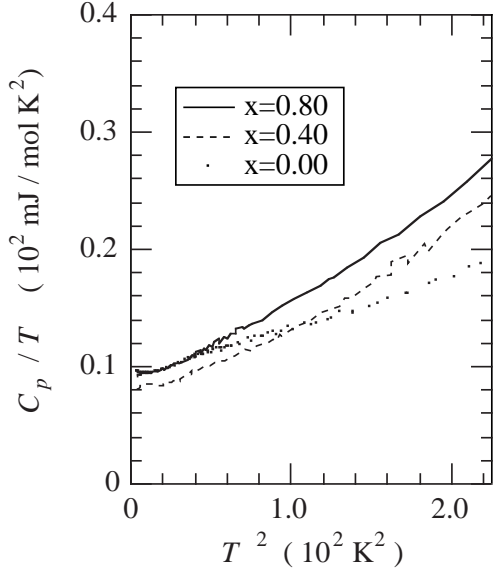,width=6.5cm}}
  \caption{%
      Specific heat of poly-crystalline \csvo\ divided by $T$ 
      plotted against $T^2$.}
  \label{f.CpT}%
\end{figure}

The measured constant pressure specific heats $C_p$ of \csvo\ below 
$\sim\!15$\,K are shown in Fig.~\ref{f.CpT}\@.
In the temperature range displayed in Fig.~\ref{f.CpT}, it is clear 
that the data do not behave 
simply as Eq.~\ref{eq.gammafit}.
Thus, we have tried to fit the data to Eq.~\ref{eq.gammafit} below 
$T^2 < 2 \times 10^2\,{\rm K^2}%
\:( T < \:\sim\!  14\,{\rm K})$, 
where the identity seems more applicable.\cite{schottky}
All the results of the least-square fits are summarized 
in Table~\ref{t.CSVOgamma}.
\begin{table*}[ht]
  \centering
  \caption{%
     Fitted parameters for the specific heat and deduced 
     effective mass \meff\ of \csvo\@.  
     $\gamma^{LDA}$ has been calculated from $D(E_F)$ obtained by the 
     LDA band calculation.\protect\cite{CVOBandCalc,DefOfDOS}  
     $m^\ast/m_b$ is defined as the ratio between $\gamma$ and 
     $\gamma^{LDA}$\@.
  \label{t.CSVOgamma}}
  \begin{tabular}{ccccc}%
    \hline%
      x    & $\Theta$\,(K) & $\gamma$\,(mJ mol^{-1}K^{-2})%
      & $\gamma^{LDA}$\, (mJ mol^{-1}K^{-2}) & $m^\ast/m_b$ \\ %
    \hline\hline%
     0.00  & $368.0$  & $9.248$ & $4.849$ & $1.907$\\ %
    \hline%
     0.20  & $348.3$  & $7.554$ & $4.706$ & $1.605$\\ %
    \hline%
     0.40  & $320.6$  & $7.123$ & $4.563$ & $1.561$\\ %
    \hline%
     0.80  & $300.0$  & $8.239$ & $4.276$ & $1.927$\\ %
    \hline%
     1.00  & $322.4$  & $8.182$ & $4.133$ & $1.980$\\ %
    \hline%
  \end{tabular}
\end{table*}
We find that the value of $\gamma$, even in \cvo\ ($x=0$), is 
still not so enhanced as we discuss below $(m^{\ast}/m_{b} \simeq 2)$, 
although these values are comparable to the La_{1-x}Sr_xTiO_3 
system in the La rich phase except for $x<0.05$\@.
(The $\gamma$ values in the \csvo\ system are much larger than 
that of the less correlated sodium metal $\sim\!1$\,mJ mol^{-1}K^{-2}.)
The obtained Debye temperatures $\Theta$ are comparable to 
$\Theta\sim300$\,K deduced from the temperature of the phonon-drag
peak of the Seebeck coefficient of \cvo,\cite{seebeck} 
substantiating the result of our least-square fit to the specific heat 
data. 

\begin{figure}[hb]
   \centerline{\psfig{file=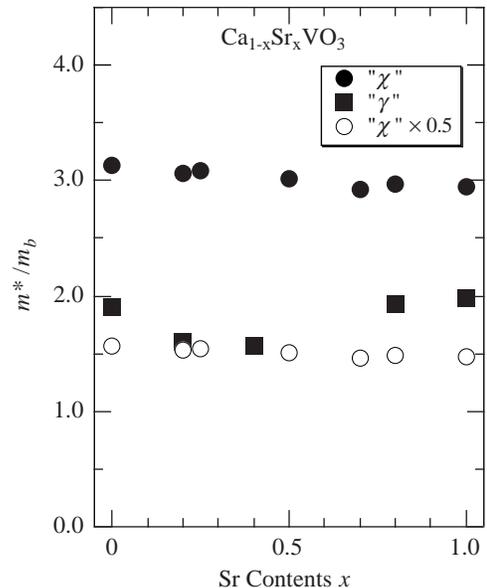,width=6.5cm}}
   \caption{%
      Ratio of the effective mass and band mass obtained by comparing the 
      observed $\gamma$ with the calculated $\gamma^{LDA}$ (filled 
      squares, denoted by ``$\gamma$'').  
      ``$\chi$'' stands for the effective mass deduced from $\chi_P$ 
      (filled circles, denoted by ``$\chi$'') and the half values 
      of ``$\chi$'' (open circles) also plotted for comparison.}
  \label{f.CSVOgammaMvsx}%
\end{figure}

The effective masses compared to the band
masses $m^\ast/m_b$ are defined as the ratios of the observed $\gamma$ to
the $\gamma^{LDA}$\@. 
We plot the values of $m^\ast/m_b$ against $x$ with those deduced from 
the magnetic susceptibilities for comparison 
(Fig.~\ref{f.CSVOgammaMvsx})\@.

From Fig.~\ref{f.CSVOgammaMvsx}, we can also estimate the 
Sommerfeld-Wilson's ratio $R_W$:\cite{wilson}
\[%
R_W \equiv %
\frac{\gamma^{LDA}}{\chi_{P}^{LDA}}\frac{\chi_{P}}{\gamma}%
=\frac{\left(\frac{m^\ast}{m_b}\right)_{\chi}%
              }{%
                \left(\frac{m^\ast}{m_b}\right)_{\gamma}}%
\]
It is worthwhile emphasizing that $R_W$ is of order unity, 
implying that the electronic 
specific heat coefficient $\gamma$ is similarly enhanced to the Pauli 
paramagnetic susceptibility $\chi_P$.
Furthermore, this means it is appropriate to assume a one-to-one 
correspondence between the quasi-particle excitations of this system 
and those of a free-electron gas.

For a non-interacting Bloch-electron system, $R_{W} = 1$\@.
One of the possible reasons for $R_{W} \neq 1$ is a ferromagnetic 
fluctuation, which enters in $\chi_P$ as $\chi_P / \chi_P^{\rm o} = 
(m^\ast/m_{\rm o})S$, where $S=(1+F_{\rm o}^a)^{-1}$ is called the Stoner 
enhancement factor including the 0-th asymmetric Landau parameter 
$F_{\rm o}^a$\@.
(For an isotropic free-electron system, $R_W$ becomes unity, because 
$F_{\rm o}^a=0$\@.)
In exchange-enhanced metals, \ie, a system with ferromagnetic 
fluctuations, $S$ plays an important role and $R_{W}$ becomes fairly 
large. However, this is not the case for the \csvo\ system, since 
we have not observed any traces of 
ferromagnetic fluctuations.

The value of $R_W$ for \csvo\ deduced from our experiments is 
$1.7 \sim 2$, as illustrated in Fig.~\ref{f.CSVOgammaMvsx}, 
where the half values 
of $m^\ast/m_b$ deduced from the magnetic measurements are plotted 
for comparison.
In strongly correlated electron systems, it has been argued that
the value of $R_{W}$ becomes equal to 2
at $U/W=\infty$\@.\cite{wilson,nozieres}
Although there is a small deviation, $R_W \approx 2$ clearly 
indicates the importance of electron correlations in this system.

The deviation from $R_{W} = 2$ can be ascribed 
to the contribution of the electron-phonon interaction.
It is known that this interaction contributes a factor 
$(1+\lambda)$ to $\gamma$, but not to $\chi_P$\@.
Hence, $R_W$ is modified to become $R_W(1+\lambda)^{-1}$\@.
From Fig.~\ref{f.CSVOgammaMvsx}, we can approximately estimate 
$\lambda < \sim\!0.3$, so that the electron-phonon interaction in this 
system is fairly small.
Furthermore, it should be noted that, for the higher orbital 
degeneracies, $R_W$ decreases towards unity in the limit of large 
orbital degeneracy.\cite{yoshimori}
Since the degeneracies of the $t_{2g}$ orbitals of the vanadium $3d$ 
electrons are not completely released in this \csvo\ system, $R_{W}$
is not necessary to be equal to 2\@.
However, there are experimental errorbars for the estimation of 
$\gamma$ values, the above argument needs to be further investigated.

Thus, we can conclude that $\sim\!1.7 < R_{W} <\:\sim\!2$ 
implies that electron
correlations are strong in this system.
Then, the question arises why is the enhancement of the effective mass 
so moderate, despite the presence of such large electron correlations?

The effective mass of a quasi-particle at the Fermi energy \ef\ is 
defined in general as; 
\begin{equation}
m^{\ast}=\left(\left.\frac{1}{\hbar^2\vec{k}}%
  \frac{d\varepsilon_k}{d\vec{k}}\right|_{\vec{k}=\vec{k}_{\rm F}}%
  \right)^{-1}\:,
\label{eq.DefEffMass}
\end{equation}
where $\varepsilon_k$ is quasi-particle energy that is given as a 
solution $\omega=\varepsilon_k$ of the equation
\begin{equation}
 \omega=\varepsilon_k^{\rm o}+{\rm Re}\Sigma(\vec{k},\omega)\:,%
 \label{eq.EpsEq}
\end{equation}
where $\Sigma(\vec{k},\omega)$ is the self-energy of the system 
in which all of the interaction effects are contained.
$\varepsilon_k^{\rm o}$ corresponds to the energy of a non-interacting 
Bloch electron.  However, in this study, we regard $\varepsilon_k^{\rm 
o}$ as the energy dispersion of a single-electron band obtained by the 
LDA band-calculation.
Thereby, $\varepsilon_k^{\rm o}$ gives a band mass \mb:
\begin{equation}
m_b=\left(\left.\frac{1}{\hbar^2\vec{k}}%
               \frac{d\varepsilon_k^{\rm o}}{d\vec{k}}\right|_{\vec{k}%
               =\vec{k}_{\rm F}}%
         \right)^{-1}\:.
\label{eq.DefBandMass}
\end{equation}
As is apparent from these definitions (Eqs.~\ref{eq.DefEffMass} and 
\ref{eq.DefBandMass}), \meff\ and \mb\ are given as tensors, but we 
assume here that the Fermi surface is isotropic and therefore \meff\ 
is nothing but a scalar quantity.
Using Eqs.~\ref{eq.DefEffMass}, \ref{eq.EpsEq} and 
\ref{eq.DefBandMass}, we deduce that the effective mass is given 
by the following expression:
\begin{eqnarray}
\frac{m^{\ast}}{m_b}%
 & = &%
  \frac{%
         \left|\left.\frac{d\varepsilon_k^{\rm o}}{d\vec{k}}%
               \right|_{\vec{k}=\vec{k}_{\rm F}}%
         \right|%
      }{%
         \left|\left.\frac{d\varepsilon_k  }{d\vec{k}}%
                \right|_{\vec{k}=\vec{k}_{\rm F}}%
         \right|%
       } \nonumber \\ %
 & = & \left( 1 - \left.%
        \frac{%
               \partial {\rm Re}\Sigma(\vec{k},\omega)%
            }{%
               \partial\omega%
             }\right|_{\omega=E_F}%
       \right) \nonumber \\ %
 & & \mbox{} \times%
 \frac{%
    \left|\left.\frac{d\varepsilon_k^{\rm o}}{d\vec{k}}%
           \right|_{\vec{k}=\vec{k}_{\rm F}}%
    \right|%
     }{%
    \left|\left.\frac{d\varepsilon_k^{\rm o}}{d\vec{k}}%
           \right|_{\vec{k}=\vec{k}_{\rm F}} + \left.%
         \frac{%
                 \partial {\rm Re}\Sigma(\vec{k},\omega)%
             }{%
                 \partial\vec{k}%
              }\right|_{\vec{k}=\vec{k}_{\rm F}}%
    \right|%
      } \nonumber \\ %
& \equiv & \frac{m_\omega}{m_b} \times \frac{m_k}{m_b} \nonumber \:,
\end{eqnarray}
where $m_\omega$ is called ``$\omega$-mass'' and $m_k$ is called 
``$k$-mass''\@.\cite{kmass}
If we consider only the on-site Coulomb interaction as the origin of 
the electron correlation
and average out the fluctuation of the neighboring sites as in the 
limit of large lattice connectivity,\cite{LISA}
the self-energy depends only  on the 
quasi-particle energy $\omega$, \ie, 
$\Sigma(\vec{k},\omega)\equiv\Sigma(\omega)$\@.
Then, the effective mass becomes (since $m_{k}/m_{b}=1$):
\[
\frac{m^{\ast}}{m_b} = 
        \left(%
           1 - \left.%
                  \frac{%
                     \partial {\rm Re}\Sigma(\omega)%
                       }{%
                     \partial\omega%
                       }%
                \right|_{\omega=E_F}%
        \right) \equiv Z^{-1}\:,
\]
where $Z$ is the quasi-particle weight.
Therefore, a critical enhancement of the effective mass due to 
strong electron correlations ($Z \rightarrow 0$) is inevitable at the 
MIT point.

However, in general, since the electron correlation is not necessarily 
confined to each atomic site, we may need to take into account the 
effect of the non-local Coulomb interaction, \ie, the self-energy 
should have a momentum-dependence.\cite{commentsigmak}
Especially in the dynamical mean-field approach to the Mott 
transition,\cite{LISA} the non-locality of the exchange interaction is 
not treated; therefore, we should introduce $\Sigma(\vec{k}_{\rm 
F},E_F)$, which is considerably different from zero.\cite{morikawa}
The screening of the Coulomb potential can reduce 
$\Sigma(\vec{k}_{\rm F},E_F)$, because in well-screened 
systems such as  conventional metals, the Coulomb potential is no 
longer long-range and the non-locality of the exchange interaction is 
small.
However, in most of the perovskite-type TM oxides, the carrier density 
is fairly small, and this effect will be more significant in the 
vicinity of the MIT point.

We must assume therefore that, near the MIT point, the $\omega$-mass 
increases significantly reflecting $Z \rightarrow 0$; on the other 
hand, due to poor-screening, the contribution of the 
momentum-dependent self-energy becomes significant, resulting in a 
decrease of the $k$-mass.
Thus, the critical enhancement of the effective mass, which is a 
product of the $\omega$-mass and the $k$-mass, can be suppressed
in some conditions.\cite{inouePRL}
This is not only a plausible idea of explaining the behavior of the 
effective mass but also a model which provides a desirable picture of the 
reduction of the spectral intensity at the Fermi energy observed in 
the photoemission and inverse-photoemission 
spectroscopies.\cite{inouePRL,morikawa,fujimorikmass}

In summary, in order to obtain a comprehensive understanding of the 
metallic state near the Mott transition, we should note that the 
momentum-dependence of the self-energy plays an important role in this 
region.

\subsection{Electric resistivity}
\label{sec.resistivity}
The electric resistivities collected on both heating and cooling 
cycles between 350\,K and 4\,K show no difference within the 
experimental accuracy.
In some cases, the resistivity shows clear dependence on the 
crystallographic alignments of the experiment; \ie, 
when the measuring current flows parallel to the \adir-axis of the 
pseudocubic perovskite, the resistivity is different from that when the 
current flows along the \abdir-axis.
This anisotropy, however, is not 
temperature-dependent.
Whenever we observe such anisotropy, we normalize each data set to the 
residual resistivity $\rho_{\rm o}$, and the resulting curves fit each 
other completely.

The scaling factor $c \equiv \rho^{[110]}(T)/\rho^{[100]}(T)$, 
varies from $\sim\!1.1$ to $\sim\!1.5$.
However there 
seems to be neither a systematic relation between $c$ and the Sr 
content $x$, nor consistency among the different sets of 
the measurements for the samples with the same value of $x$.

\begin{figure}[ht]
  \centerline{\psfig{file=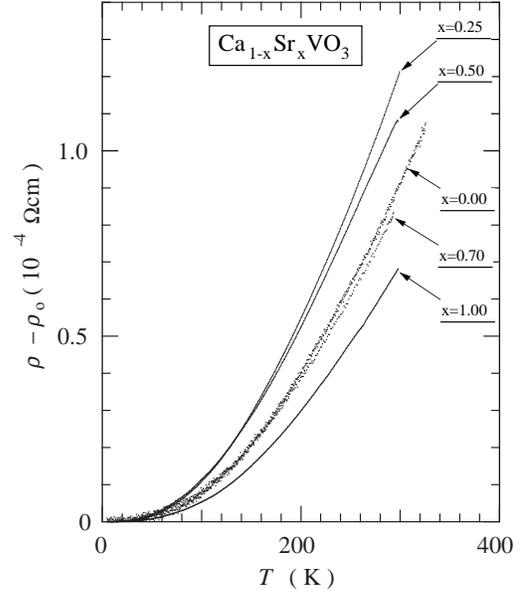,width=7.0cm}}
  \caption{%
       Electric resistivities of the \csvo\ single crystals for $x=0.00, 
       0.25, 0.50, 0.70, 1.00$\@.
       For each data set,
       the minimum resistivity at $\sim\!4$\,K has been subtracted as the 
       residual resistivity $\rho_{\rm o}$\@.}
  \label{f.CSVOrhoALL}%
\end{figure}

Thus, we consider that the observed anisotropy is not due to 
any particular feature of the electronic structure of the system.
Similar behavior has been reported in the resistivity of the 
single crystal CoSi_2 with cubic C1 structure\cite{CoSiResistivity}, 
and also high-purity cubic Al single crystal.\cite{AlResistivity}
In the former material, it was pointed out that the anisotropy can be 
attributed to an extrinsic origin, {\it e.g.}, point defects and/or 
dislocations that appeared during crystal growth, though no 
trace of such defects has yet been observed.\cite{CoSiResistivity}
It has been argued that, in the case of the Al single crystal, a 
model calculation for $\langle211\rangle$ dislocations predicts an 
anisotropy of electrical resistivity compatible with 
experiment.\cite{AlResistivityCalc}
Therefore, we suggest that the anisotropy in our resistivity 
measurements may also be caused by the presence of a small
amount of defects and/or 
dislocations.\cite{commentResistivity}
Despite this undesirable artifact, if we make the size of the 
rectangular parallelepiped as small as $2\times0.5\times0.3$\,mm^3, 
the absolute values of the electric resistivity data can be reproduced 
within the $\pm\!15$\,\% error bar (the temperature dependence is 
completely reproducible as mentioned above) irrespective of the 
direction of the measuring current.
With all these considerations, the data were collected as shown in 
Fig.~\ref{f.CSVOrhoALL}\@.

At first sight, all the data seem to be well expressed by the 
relation $\rho=\rho_{\rm o} + AT^2$ for the measured temperature range. 
However, when we try to fit the observed resistivity using this 
expression, we cannot fit the data over the entire temperature range 
from 4\,K to 350\,K using a single value of the coefficient $A$\@.
Therefore we assume that the resistivity is expressed by 
$\rho_{\rm o}+AT^2$ plus an {\it additional term}.

Firstly, we consider here that the $AT^2$ term is due to electron-phonon 
scattering.
It has been suggested, especially in strongly-coupled 
superconductors, that the $AT^2$ term is due to the 
breakdown of the momentum-conservation law in the electron-phonon 
scattering process.\cite{RhoElectronPhonon}
Here the coefficient $A$ 
is shown to be related to both the residual resistivity $\rho_{\rm o}$ and 
the Debye temperature $\Theta$:
\begin{equation}%
  A=\alpha \times \frac{\rho_{\rm o}}{\Theta^2}
  \label{eq.electronphonon}
\end{equation}
with $\alpha$ varying from $\sim\!0.01$ to $\sim\!0.1$\@.  
However, the $A$ values in the \csvo\ system, which are roughly 
estimated as $\sim\!1 \times 10^{-9}\,\Omega {\rm cm/K^2}$, are three 
orders larger than
\begin{eqnarray*}%
 \alpha \times \frac{\rho_{\rm o}}{\Theta^2}&=& %
 \alpha \times%
 \frac{\sim\!1 \times 10^{-5}\,(\Omega {\rm cm})}{%
 (\sim\!5 \times 10^2)^2\,({\rm K^2})} \\ %
  &=& \sim\!4 \times 10^{-12}\,(\Omega {\rm cm/K^2})\:,%
\end{eqnarray*}%
even if we assume the largest value of $\alpha\sim\!0.1$\@. 
(Here we used the Debye temperature estimated from the specific heat 
measurement.)
Furthermore, Gurvitch has discussed that a strong 
electron-phonon interaction is insufficient for the $T^2$ law; 
the simultaneous presence of strong coupling and disorder is 
also necessary.\cite{gurvitch}
He has also pointed out that, in some cases, 
Eq.~\ref{eq.electronphonon} is not applicable; \ie, there is an empirical 
condition for the appearance of the $T^2$ law:
\[%
(\lambda-0.7)\times\rho_{\rm o}>\;\sim\!13\,{\rm (\mu\Omega cm)}\:,
\]
where $\lambda$ is the electron-phonon coupling constant. 
In the \csvo\ system, however, $\lambda$ is at largest $\sim\!0.3$ as 
discussed in Sec.~\ref{sec.gamma}, and $\rho_{\rm o}$ is 
$\sim\!1\times 10^{-5}\,{\rm \Omega cm}$.
Hence, the last formula is not 
satisfied.
(In the first place, even the value of $\lambda$ 
is smaller than $0.7$.)
Following these arguments, it appears unlikely that
the $T^{2}$-dependent resistivity in \csvo\ arises from
electron phonon scattering. 
However, this kind of contribution to the $T^{2}$ term is not completely 
neglected and will be discussed again below. 

\begin{figure}[ht]
  \centerline{\psfig{file=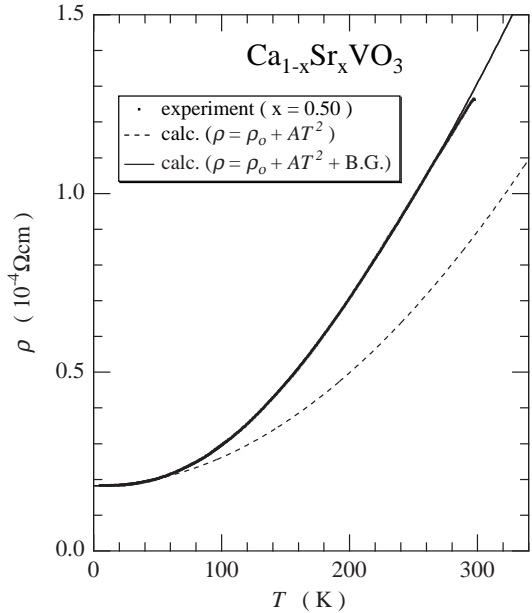,width=7.0cm}}
  \caption{%
      Electric resistivity of Ca_{0.5}Sr_{0.5}VO_3 against temperature $T$ 
      (dots)\@.
      The solid line represents Eq.~\protect\ref{rhofit}, while the broken 
      line represents Eq.~\protect\ref{rhofit} without the 
      Bloch-Gr\"{u}neisen term.}
  \label{f.x050rhofit}%
\end{figure}

An alternative and more likely origin of the $T^2$ term is 
electron-electron scattering in the presence of the umklapp process.  
Let us consider here the resistivity 
as modeled by a three-component expression of the form:
\[
  \rho=\rho_{\rm o}+\rho_{\hbox{\scriptsize e-e}}(T)%
  +\rho_{\hbox{\scriptsize e-ph}}(T)%
\]
where $\rho_{\rm o}$ is a temperature-independent background 
contribution due to static disorder, $\rho_{\hbox{\scriptsize e-e}}(T) 
\equiv AT^2$ is the electron-electron scattering.
By the line-shape analysis such as is shown in Fig.~\ref{f.x050rhofit}, we 
found that the third term $\rho_{\hbox{\scriptsize e-ph}}(T)$ is well 
represented by the 
classical Bloch-Gr\"{u}neisen formula for electron-phonon scattering 
with $n=5$, 
developed for an isotropic Fermi surface and a simple phonon spectrum:
\begin{equation}
\rho = \rho_{\rm o} + AT^2 
 + \frac{4 \kappa T^n}{\Theta^6}
 \int_{0}^{\Theta/T}\frac{{\rm e}^z z^n dz}{({\rm e}^z - 1)^2}\mbox{  }(n=5)
  \label{rhofit}
\end{equation} 

\begin{figure}[ht]
  \centerline{\psfig{file=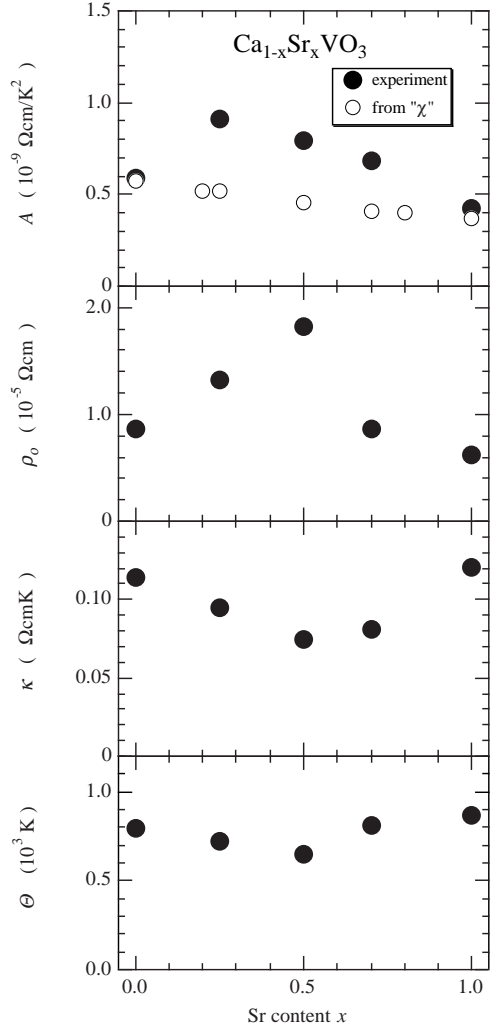,width=6.5cm}}
  \caption{%
      Fitted parameters for the electric resistivity: $A$, 
      $\rho_{\rm o}$, $\kappa$ and $\Theta$ in Eq.~\protect\ref{rhofit}, plotted 
      against Sr content $x$\@. The $A$ values deduced from the Pauli
      paramagnetic susceptibility under the assumption of $R_{W}=2$ and
      the Kadowaki-Woods ratio is 
	  $1.0\times10^{-5}$\,$\mu\Omega$\,cm\,mol^{2}\,K^{2}\,(mJ)^{-2} 
	  are also plotted for comparison (top)\@.}
  \label{f.rhoparametersSrx}%
\end{figure}

We have done a least-square fit to all the data using Eq.~\ref{rhofit}
and the obtained parameters are summarized in Table~\ref{t.BGfit} and 
also in Fig.~\ref{f.rhoparametersSrx}\@.
If we accept $\sim\!\pm15$\,\% error bar, we can conclude that
each of the fitted parameters shows systematic behavior as a function
of Sr content $x$\@.
\begin{table*}[ht]
  \centering
  \caption{%
    Fitted parameters for the electric resistivity of \csvo\ with
    Eq.~\protect\ref{rhofit}\@.
  \label{t.BGfit}}
  \begin{tabular}{ccccc}%
    \hline%
       x   & $\rho_{\rm o}$\,($\Omega$cm) & $A$\,(($\Omega$cm/K^2)  &%
          $\kappa$\,($\Omega$cm K) & $\Theta$\, (K) \\ %
    \hline\hline%
      0.00 & $8.668 \times 10^{-6}$ & $5.911 \times 10^{-10}$ &%
          $0.114$                 & $793.5$        \\ %
    \hline%
      0.25 & $1.319 \times 10^{-5}$ & $9.118 \times 10^{-10}$ &%
          $9.476 \times 10^{-2}$  & $722.2$        \\ %
    \hline%
      0.50 & $1.827 \times 10^{-5}$ & $7.900 \times 10^{-10}$ &%
          $7.441 \times 10^{-2}$  & $647.3$        \\ %
    \hline%
      0.70 & $8.656 \times 10^{-6}$ & $6.796 \times 10^{-10}$ &%
          $8.073 \times 10^{-2}$  & $811.5$        \\ %
    \hline%
      1.00 & $6.205 \times 10^{-6}$ & $4.208 \times 10^{-10}$ &%
          $0.121$                 & $866.3$        \\ %
    \hline%
  \end{tabular}
\end{table*}

$\rho_{\rm o}$ shows a maximum at $x=0.5$.
This reflects that the system has the maximum amount of randomness at 
that composition.

We note that our effective transport Debye temperature $\Theta \sim 
700$\,K does not sound physical.
However it is not necessary for the transport $\Theta$ to be 
equal to the thermodynamic value $\Theta \sim 350$\,K obtained from 
the specific heat measurements.
This is because the transport Debye temperature involves only the 
acoustic modes that interact with the electrons, whereas the 
thermodynamic Debye temperature considers 
all types of phonons.\cite{bucher}

The values of $\kappa$ and $\Theta$ show a minimum at $x=0.5$, 
indicating that the lattice becomes softest at this composition.
The electron-phonon coupling constant $\lambda$ is related to both 
$\kappa$ and $\Theta$ as follows:
\[%
 \lambda \propto \frac{\omega_p^2}{\Theta^2}\,\kappa\:,
\]
where $\omega_p$ is the plasma frequency of the conduction electrons.
Makino {\it et al}.\ reports in the following paper\cite{makinoCSVO} 
that the variation of $\omega_p$ in going from \svo\ to \cvo\ is 
systematic but very small, and the variation of $\lambda$ inferred 
from the Sommerfeld-Wilson's ratio $R_W$ is also small.
Thus we can roughly estimate that $\kappa \sim \Theta^2$\@.
This is consistent with the behaviors shown 
in Fig.~\ref{f.rhoparametersSrx}\@.

It is surprising that the contribution of the electron-electron 
scattering, which is in general dominant at very low temperature, is 
significantly large even at room temperature.
In Ca_{0.5}Sr_{0.5}VO_3,
\[%
\rho_{\hbox{\scriptsize e-e}}(300\,{\rm K}) :%
 \rho_{\hbox{\scriptsize e-ph}}(300\,{\rm K}) \sim 2 : 1\: .%
\]%
This is further evidence that the electron correlations are 
significantly large in this system.
The coefficient $A$ should, then, reflect the enhancement of the 
effective mass of the quasi-particles due to this electron 
correlation.
The resistivity due to electron-electron (\ie, 
quasi-particle\,$-$\,quasi-particle) scattering can be crudely but 
quantitatively expressed as follows:
\begin{eqnarray*}
\rho_{\hbox{\scriptsize e-e}}(T)&=& \frac{m_{b}}{ne^2\tau} \\ %
 &=& \frac{m^{\ast} v_F}{e^2 n^{2/3}}\left(\frac{k_B T}{E_F}\right)^2 \\ %
 &=& \frac{4k_B^2 \left.m^\ast\right.^2}{\hbar^3 e^2 n^{2/3}k_F^3} T^2 \\ %
 &\equiv& A T^2\:,
\end{eqnarray*}
where $v_F = \hbar k_F / m^{\ast}$, $E_F = \hbar^2 k_F^2/(2 m^{\ast})$, and 
we have assumed the scattering time $\tau$ is equal to the life time of the 
thermally activated quasiparticle 
$\tau^{-1}=v_{F}n^{1/3}E_{F}^{-1} Z^{-1}{\rm Im}\Sigma%
\simeq v_Fn^{1/3}(m^{\ast}/m_{b})(k_B T/E_{F})^{2}$\@.\cite{commentAT} 
Hence, the coefficient $A$ is proportional to the quadratic of the 
effective mass \meff\@.
The obtained value of $A$ increases systematically in going from $x=1$ 
to $x=0.25$; this may correspond to the increase of \meff. 
However there 
seems to be a rapid decrease between $x=0.25$ and $x=0$ which is not 
consistent with the behavior of \meff\@.
This is not obviously explained by the considerably large error
arising from the quality of the sample ($\sim\!\pm15$\,\%).

In Fig.\ref{f.rhoparametersSrx} (top), 
we also plot the value of $A$ deduced from the Pauli paramagnetic
susceptibility $\chi_{P}$ under a few reasonable assumptions:
we use a Sommerfeld-Wilson's ratio $R_{W}=2$ to estimate the electronic 
specific heat $\gamma$ from $\chi_{P}$, since $R_{W}$ is almost equal 
to 2
in this system.
The obtained $\gamma$ corresponds to the electronic 
specific heat which is  not affected by the electron-phonon 
interaction.
Then we use the Kadowaki-Woods ratio $A/\gamma^{2}$,
which is a measure of the electron correlation,
of the same value 
as that of the heavy fermion systems,\cite{kadowakiwoods}
\ie\ $A/\gamma^{2}=%
1.0\times10^{-5}$\,$\mu\Omega$\,cm\,mol^{2}\,K^{2}\,(mJ)^{-2},
and deduced the value of $A$.
The resulting $A$ values are compared to the experimentally observed 
values (Fig.~\ref{f.rhoparametersSrx}, top)\@.

Since the above assumptions for the Sommerfeld-Wilson's ratio and the 
Kadowaki-Woods ratio in this system seem to be fairly appropriate, 
we can safely say
that the $AT^{2}$ term in the resistivity of the end members 
\cvo\ $(x=0)$ and \svo\ $(x=1)$ is 
attributed to only the electron-electron scattering. For the 
other solid-solutions $(0<x<1)$, there must be other contributions
to the $AT^{2}$ term.
The most probable candidate of this additional contribution of the 
$T^{2}$ term is the interference between the elastic electron 
scattering and the electron-phonon scattering, which has been recently 
investigated by Ptitsina {\it et al}\@.\cite{ptitsina}
This effect must be proportional to the residual resistivity, and our data 
seems to support the scenario.

On the other hand, however, we know that the Kadowaki-Woods ratio 
is not necessarily equal to the above value. 
Then, there are several other reasons to be considered for the observed 
$x$-dependence of $A$\@.
We must consider a possible contribution from the 
modification of the Fermi surface.
Since $A$ is not only proportional to 
$\left.m^{\ast}\right.^{2}$ but also to $\left.k_F\right.^{-3}$, 
a variation of 
the shape of the Fermi surface due to the orthorhombic 
distortion may lead to changes in $A$.
In passing, as we apply pressure to \cvo, the value of $A$ tends to 
decrease,\cite{seebeck,igapressure} though it has not yet been 
determined how the lattice constants change under pressure.
We should also remark that samples around 
$x=0.25$ have a tendency to show the smallest 
spectral weight at the Fermi 
energy as observed in the recent studies of the photoemission 
spectroscopy\cite{inoueCSVOPES} and the inverse photoemission 
spectroscopy of the \csvo\ single crystals.\cite{ddsarma}
This might be related to the hybridization between the V $3d$ $t_{2g}$ 
orbitals and the O $2p$ $\sigma$ orbitals.
Okimoto {\it et al}.\ calculated a distortion-induced admixture between 
those orbitals,\cite{okimoto} and Lombardo {\it et al}.\ discussed a 
possible scenario involving spectral weight transfer in the \csvo\ system 
due to charge transfer.\cite{lombardo}
This distortion-induced charge transfer may explain the strange revival
of the quasi-particle weight in the region close to \cvo, 
and this hybridization may also explain 
the $x$-dependence of the value of $A$,
although this 
should be consistent with the monotonic increase of the effective mass 
toward \cvo\ as estimated from the Pauli paramagnetism.
Finally, the contribution of the momentum-dependence of 
the self-energy, which 
becomes significant in the region closer to 
\cvo,\cite{inouePRL,morikawa} is also an intriguing candidate to 
explain the behavior of $A$.
Apparently, with a large 
momentum-dependence of the self-energy, it is no longer necessary that 
$A$ is proportional to $\left.m^{\ast}\right.^{2}$\@.
These issues will be clarified by further investigations.

\section{SUMMARY AND CONCLUDING REMARKS}
We have succeeded to prepare single crystals of the metallic alloy 
system \csvo\ for the first time, as far as we are aware.
The system has nominally one $3d$ electron per vanadium ion; as we 
substitute a \Ca\ ion for a \Sr\ ion, the band width $W$ decreases due 
to the buckling of the V-O-V bond angle from $\sim\!180^\circ$ for \svo\ 
to $\sim\!160^\circ$ for \cvo, which is almost equal to the analogous 
$3d^{1}$ insulator LaTiO_3\@.
Thereby, it is reasonable to consider that \cvo\ is close
to the boundary of MIT. 
The Sommerfeld-Wilson's ratio $R_{W}\simeq2$, the Kadowaki-Woods ratio 
$A/\gamma^{2}$ lies in the same region as the heavy Fermion compounds, and 
there is a large contribution from electron-electron scattering to 
the resistivity even at room temperature.
These features are considered to provide strong 
evidence of the large electron correlations in this system.
However, the effective masses obtained by the thermodynamic $(\gamma)$ 
and magnetic $(\chi_P)$ measurements show only a moderate increase in 
going from \svo\ to \cvo, instead of the diverging behaviors expected 
from the Brinkmann-Rice picture.

The elaborate band-width control in \csvo\ 
has elucidated that the mass-enhancement due to the reduction 
of the band width is not so large, even though the system shows some 
fingerprints of a large electron correlation.
Accordingly, we suggest that these seemingly contradicting metallic 
properties observed in this system can be explained by considering the 
effect of non-local electron correlations, \ie, the momentum-dependent 
self-energy.

\section{ACKNOWLEDGEMENT}
We would like to thank A.  Fujimori, M.  J.  Rozenberg, A.  
Yoshimori, H.  Bando, N.  Shirakawa, and I.  Hase for many fruitful 
and stimulating discussions.
We also thank Y. Ueda for showing the data of
the crystal structure of \cvo\ before the publication.
Finally, we gratefully acknowledge many experimental supports and 
suggestions, at various stages of this work, by Y.  Nishihara, K.  
Oka, F.  Iga, H.  Kawanaka, A.  Fukushima, T.  Ito, Y.  Kodama and 
other members of Electron Physics group in Electrotechnical 
Laboratory.


\begin{thebibliography}{99}
\bibitem{commentOtherPossibility}
The GdFeO_{3} type lattice distortion possibly causes anisotropy
and affects electron-phonon interactions. However, in this system,
these contributions are not predominant for the metallic properties, 
as revealed by our 
measurements  and the following discussion. This could also
justify the metallic \csvo\ alloy is an ideal compounds.
\bibitem{gammasingle} Very recently, we have obtained preliminary 
specific heat data for single crystals, which are almost similar
to those for the poly crystals.
\bibitem{DefOfDOS}
For $D(E_F)$, we have used
the LDA DOS at \ef\ 
for \cvo\ and \svo, and have simply made a linear interpolation for the 
intermediate Sr/Ca stoichiometries.
\bibitem{schottky} 
A bump which appears around 2\,K in the specific heat can be
subtracted by assuming Schottky contribution of $\sim\!0.2$\,\% 
impurities. This amount of impurities is consistent with that deduced
from the magnetic Curie-Weiss contribution.
\bibitem{commentResistivity}
When the rate of crystal growth is fairly large, it is hard to 
explain this process on a plane by plane basis.  However, if 
the crystal contains a screw dislocation, it is never necessary to 
nucleate a new plane, as the local planar structure can wind 
endlessly about the screw dislocation.  This kind of process might be 
realized in some cases during our crystal growth.
\bibitem{commentAT}
In strongly correlated electron systems, the electron-electron 
scattering is the dominant process to induce the decoherence of the
wavefunction, which thus determines the 
relaxation time $\tau$ under the presence of the umklapp process. 
Then, the 
quasi-particle\,$-$\,quasi-particle scattering time is given by 
$\tau^{-1} \propto Z^{-1}{\rm Im}\Sigma$, 
where $Z$ is the quasi-particle weight and 
$\Sigma$ is the self-energy.
\end{thebibliography}
\end{document}